\documentclass[twocolumn,english,aps,pra,superscriptaddress,showpacs,tightenlines]{revtex4-1}
\usepackage{amsmath}
\usepackage{graphicx}
\usepackage{amssymb}
\usepackage{CJK}
\usepackage{color}

\begin{document}

\begin{CJK*}{GBK}{song}

\title{tunable single-photon frequency converter in a waveguide with a giant V-type atom}
\author{Hongzheng \surname{Wu} }
\affiliation{Key Laboratory of Low-Dimension Quantum Structures and Quantum Control of Ministry of Education, Key Laboratory for Matter Microstructure and Function of Hunan Province, Synergetic Innovation Center for Quantum Effects and Applications, Xiangjiang-Laboratory and Department of Physics, Hunan Normal University, Changsha 410081, China}
\author{Ge \surname{Sun} }
\affiliation{Key Laboratory of Low-Dimension Quantum Structures and Quantum Control of Ministry of Education, Key Laboratory for Matter Microstructure and Function of Hunan Province, Synergetic Innovation Center for Quantum Effects and Applications, Xiangjiang-Laboratory and Department of Physics, Hunan Normal University, Changsha 410081, China}
\affiliation{Institute of Interdisciplinary Studies, Hunan Normal University, Changsha, 410081, China}
\author{Jing \surname{Lu} }
\affiliation{Key Laboratory of Low-Dimension Quantum Structures and Quantum Control of Ministry of Education, Key Laboratory for Matter Microstructure and Function of Hunan Province, Synergetic Innovation Center for Quantum Effects and Applications, Xiangjiang-Laboratory and Department of Physics, Hunan Normal University, Changsha 410081, China}
\affiliation{Institute of Interdisciplinary Studies, Hunan Normal University, Changsha, 410081, China}
\author{Lan \surname{Zhou} }
\thanks{Corresponding author}
\email{zhoulan@hunnu.edu.cn}
\affiliation{Key Laboratory of Low-Dimension Quantum Structures and Quantum Control of Ministry of Education, Key Laboratory for Matter Microstructure and Function of Hunan Province, Synergetic Innovation Center for Quantum Effects and Applications, Xiangjiang-Laboratory and Department of Physics, Hunan Normal University, Changsha 410081, China}
\affiliation{Institute of Interdisciplinary Studies, Hunan Normal University, Changsha, 410081, China}

\begin{abstract}
We study the single-photon scattering in a one-dimensional (1D) waveguide coupled to one transition of a $V$-type
giant atom (GA), whose other transition is coherently driven by an classical field. The inelastic scattering of
single photons by the GA realizes the single-photon frequency conversion. By applying the Lippmann-Schwinger
equation, the scattering coefficients for single photons incident from different directions are obtained,
which present different scattering spectra in the Markovian and the non-Markovian regimes. The conversion contrast
characterizing the nonreciprocity is also analyzed in both regimes. It is found that the probability of the frequency
up- or down-conversion vanishes as long as the emission from either transition pathways for single photons is suppressed,
but it is enhanced and even reach unity by introducing the nonreciprocity. It is the quantum self-interference induced
by the scale of this two-legged GA and the phase difference between the GA-waveguide couplings that tune the probability
of the frequency up- or down-conversion.

\end{abstract}
\maketitle

\end{CJK*}\narrowtext

\section{Introduction}

Photons are ideal carriers of quantum information, as they interact only
weakly with the environment, and their processes are fast and efficient. The
practical absence of photon-photon interactions at the single-photon level
in a vacuum makes it necessary to rely on the coupling of quantum emitters
(QEs) and photons in order to attain nonlinearities at the level of
individual photons, unfortunately, the strength of the single atom-photon
interaction is intrinsically weak in free space. Nevertheless, the coupling
can be substantially increased by placing the QEs in either high-finesse
cavities\cite{436N87,118PRL133604} or waveguides with strong-field
confinement~\cite{6NP93,6NC8655}. In a one dimensional (1D) waveguide, the
electromagnetic field is confined spatially in the transverse direction and
maintain uniform in the longitudinal direction, which forms an effective
one-dimensional (1D) continuum for propagating photons. The two ends of the
waveguide are two natural ports for introducing and extracting information,
and the photonic architecture of the waveguide coupled with QEs is
convenient to design complex optical quantum devices, such as single-photon
switches~\cite{swPRA74(06),LanPRL101,swPRA100(19),swPRA102(20),swPRA110(24)}
and routers~\cite%
{routPRL107,lanPRL111,LuPRA89,LuOE23,YangPRA98,routPRA105,routPRA94,WeiPRA89,AhumPRA99,routPRAA15}%
, diodes~\cite{diodePRB81,diodePRA98,ZhouOE30}, et.al..

With the great technological progress of superconducting circuits,
experiments on superconducting artificial QEs coupled to surface acoustic
waves~\cite{SAWSci346,SAWSNC8,SAWSPR124} or microwaves through multiple
coupling points through suitably meandering the transmission line with
wavelength-scale distance~\cite{GAWPRA103} have required going beyond the
treatment of a QE as pointlike matter coupling locally to a waveguide. Such
QEs in the literature are called ``giant atoms (GAs)''. Since the size of
GAs is comparable to or much larger than the wavelength of the
electromagnetic radiation, the dipole approximation is no longer valid. The
nonlocal GA-photon interaction leads to remarkable phenomena that are not
present in quantum optics with pointlike QEs, such as frequency-dependent
decay rate and collective Lamb shift~\cite{GAP583N,GAP103PRA,GAP90PRA} due
to the self-interference induced by the scale of a GA, non-Markovian
dynamics~\cite{NM15NP1123,NM45OL3017,NM116PRL093601,NM133PRL063603} and
tunable atom-photon bound states~\cite%
{GAB2PRR,GAB106PRL,GAB107PRA,GAB108PRA1,GAB109PRA,GAB6PRR,GAB109PRA1,GAB111PRA,GAB111PRA1}
due to the significant internal time-delay for the field to propagate across
the GA, waveguide-mediated GA-GA interaction~\cite%
{GAE130PRL,GAE120PRL,GAE105PRA,GAE109PRA,GAE111PRA,GAE111PRAY},
nonreciprocal excitation transfer between GAs~\cite{GAN103PRAY,GAN104PRAW}
in an ordinary waveguide, etc..

To interface diverse quantum nodes that operate at different wavelengths in
a quantum network, frequency conversion of photons are required. Although
frequency conversion has been explored in the large photon flux limit, the
capability of converting single photons from one frequency to another
frequency is desired at the single-photon level. Frequency converter have
been proposed for single photons by a point-like three-level atom locally
coupled to a Sagnac interferometer~\cite{QFC108PRLS}, a coupled-resonator
waveguide~\cite{QFC89PRAW}, a semi-infinite 1D transmission line~\cite%
{QFC96PRAL}. Here, we propose a single-photon frequency converter based on
the phases of the GA-waveguide coupling strengths. We notice that frequency
conversions for a 1D waveguide with a GA have been proposed in Ref.~\cite%
{QFC104PRAD,QFC3PRRD}, however, the conversion efficiency in a convetional
waveguide is at most one-half and the Markovian approximation is made in Ref~%
\cite{QFC104PRAD}, then two Sagnac interferometers are used to enhance the
conversion efficiency. A conversion efficiency close to unity is found in
Ref.~\cite{QFC3PRRD}, however chiral atom-waveguide couplings are necessary.
In this paper, we exploit the self-interference induced by the scale of a
two-legged GA to achieve single-photon frequency conversion with efficiency
close to unity in a conventional waveguide with linear light dispersion. We
apply the Lippmann-Schwinger equation for obtaining the scattering
amplitudes when the travel time of the photon between adjacent coupling
points is comparable to or larger than the inverse of the bare relaxation
rate of the GA, then study the influence of the accumulated phases that
photons travel between coupling points and the phase difference between the
two coupling points on the scattering spectra and the conversion contrast in
the Markovian and Non-Markovian regimes. The inelastic scattering of single
photons by the GA enables the frequency conversion. The conversion
probability in the system with reciprocity is at most one-half. To further
improve the conversion probability, breaking the reciprocity is necessary.
Thanks to the phase difference between GA-waveguide couplings which enables
the nonreciprocal for photon scattering, the probability of the frequency
up- or down-conversion increases, and it even reached unity when
atom-waveguide coupling are not chiral.


\section{\label{Sec:2}Model and The Scattering Coefficients}


We consider a three-level giant atom (GA) with levels $|g\rangle $,$%
|f\rangle $, $|e\rangle $ form a V-type configuration, as shown in Fig.\ref%
{fig0}. A classical driving field with frequency $\omega _{d}$ drives the
transition $|f\rangle \leftrightarrow |e\rangle $ with the Rabi frequency $%
\Omega $, and the transition $|g\rangle \leftrightarrow |e\rangle $ is
coupled at positions $\pm d/2$ to a 1D waveguide with coupling strengths $%
J_{j},j=1,2$, where couplings $J_{j}$ are general complex. The 1D waveguide
supports a continuum of electromagnetic modes, each associated with a wave
vector $k$, frequency $\omega _{k}=v|k|$ and annihilation operators $\hat{a}%
_{k}$. The total Hamiltonian for this system reads
\begin{eqnarray}
H &=&\omega _{e}|e\rangle \langle e|+(\omega _{f}-\omega _{d})|f\rangle
\langle f|+\int dk\omega _{k}\hat{a}_{k}^{\dagger }\hat{a}_{k}  \label{M-01}
\\
&&+\Omega |f\rangle \langle g|+\sum_{j=1}^{2}\int dkJ_{j}e^{\left( -1\right)
^{j}\mathrm{i}k\frac{d}{2}}\hat{a}_{k}^{\dagger }\left\vert g\right\rangle
\left\langle e\right\vert +H.c.  \notag
\end{eqnarray}%
The ground-state energy is set as the reference so that the energies of $%
|e\rangle $ and $|f\rangle $ are denoted by $\omega _{f}$ and $\omega _{e}$,
respectively. $\Omega $ is taken to be real. To write Eq.(\ref{M-01}) in the
rotating-wave approximation, it has been assumed that $\omega _{f}\simeq
\omega _{d}$ and that frequencies $\omega _{i},i=e,f,d$ are much larger than
the coupling strengths $J_{j}$ and $\Omega $.
\begin{figure}[tbp]
\includegraphics[width=8 cm,clip]{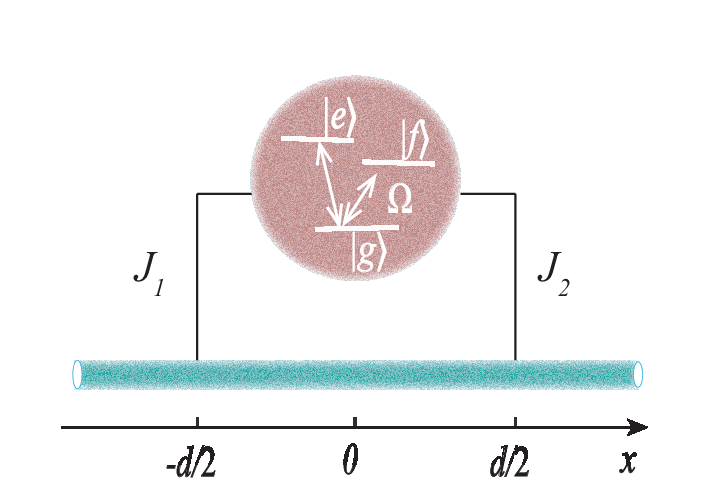}
\caption{Schematic of frequency conversion using a three-level GA with a
V-type configuration coupled to a linear waveguide. The transition $%
|e\rangle \leftrightarrow |g\rangle $ is coupled to the waveguide at
positions $x=-d/2$ and $x=d/2$, and transition $|g\rangle \leftrightarrow
|f\rangle $ is driven by a classical field with frequency $\protect\omega %
_{d}$ and Rabi frequency $\Omega $.}
\label{fig0}
\end{figure}

To find the GA's effect on a single photon incident from the waveguide, we
introduce two dressed states
\begin{subequations}
\begin{eqnarray}  \label{M-02}
\left\vert \eta_{+}\right\rangle &=& \sin\frac{\theta}{2}\left\vert
g\right\rangle +\cos\frac{\theta}{2}\left\vert f\right\rangle \\
\left\vert \eta_{-}\right\rangle &=& -\cos\frac{\theta}{2}\left\vert
g\right\rangle +\sin\frac{\theta}{2}\left\vert f\right\rangle
\end{eqnarray}
with the corresponding eigenenergies
\end{subequations}
\begin{equation}  \label{M-03}
\nu_{\pm}=\frac{(\omega_{f}-\omega_{d})\pm\sqrt{(\omega_{f}-%
\omega_{d})^{2}+4\left\vert \Omega\right\vert ^{2}}}{2},
\end{equation}
and $\tan\theta=2\Omega/(\omega_{f}-\omega_{d})$. The Hamiltonian in Eq.(\ref%
{M-01}) can be rewritten in terms of the dressed states as $H=H_0+V$
\begin{subequations}
\label{M-04}
\begin{eqnarray}
H_0 &=& \sum_{n=\pm}\nu_{n}\left\vert \eta_{n}\right\rangle \left\langle
\eta_{n}\right\vert +\omega_{e}\left\vert e\right\rangle \left\langle
e\right\vert +\int dk\omega_{k}\hat{a}_{k}^{\dagger}\hat{a}_{k} \\
V &=& \sum_{j=1}^{2}\int dkJ_{js}e^{\left( -1\right) ^{j}\mathrm{i}k\frac{d}{%
2}}\hat{a}_{k}^{\dagger}\left\vert \eta_{+}\right\rangle \left\langle
e\right\vert+H.c. \\
&&-\sum_{j=1}^{2}\int dkJ_{jc}e^{\left( -1\right)^{j}\mathrm{i}k\frac{d}{2}}%
\hat{a} _{k}^{\dagger}\left\vert \eta_{-}\right\rangle \left\langle
e\right\vert-H.c.  \notag
\end{eqnarray}
with $J_{js}=J_{j}\sin\frac{\theta}{2}$ and $J_{jc}=J_{j}\cos\frac{\theta}{2}
$. Here, $H_0$ is the free Hamiltonian for the 1D waveguide and the GA, $V$
describes the GA-waveguide interaction. When an incoming photon with
wavenumber $k$ excites the GA from its initial state $\left\vert
\eta_{-}\right\rangle$ ($\left\vert \eta_{+}\right\rangle$), the excited GA
can spontaneously decay to either state $\left\vert \eta_{-}\right\rangle$
or $\left\vert \eta_{+}\right\rangle$ and emit a photon with frequency
either unchanged or down-shifted (up-conversion). Consequently, the
three-level atom acts as a frequency converter for single photons
propagating in the 1D waveguide.

The total excitation number operator $\hat{N}=\int dk\hat{a}_{k}^{\dagger}%
\hat{a}_{k}+|e\rangle\langle e|$ commutes with the Hamiltonian $H$, states $%
\hat{a}^{\dagger}_k|0\eta_{n}\rangle, |0e\rangle$ are the eigenstates of the
number operator $\hat{N}$ in the subspace with one excitation, and they can
expand the eigenstate $\left\vert \psi_{nk}^{+}\right\rangle$ of the
Hamiltonian with a total energy $E_{nk}=v|k|+\nu_n$ as
\end{subequations}
\begin{eqnarray}  \label{M-05}
\left\vert \psi_{nk}^{+}\right\rangle =\int dq\hat{a}_{q}^{\dagger
}\left(\beta_{q}\left\vert 0\eta_{+}\right\rangle+ \chi_{q}\left\vert
0\eta_{-}\right\rangle\right) +u_{kn}\left\vert 0e\right\rangle,
\end{eqnarray}
where $u_{kn}$ is the probability amplitude of the GA in the excited state
while the field is in vacuum. $\beta_{q}$ ($\chi_{q}$) is the single-photon
field amplitude while the GA is in the state $|\eta_{+}\rangle$ ($%
|\eta_{-}\rangle$). For photons propagating in a 1D waveguide and
interacting with a GA scatterer, eigenstate $\left\vert
\psi_{nk}^{+}\right\rangle$ also satisfies the Lippman-Schwinger equation~%
\cite{LSequbook}
\begin{equation}  \label{M-06}
\left\vert \psi_{nk}^{+}\right\rangle=\hat{a}_{k}^{\dagger}\left\vert
0\eta_{n}\right\rangle +\frac{1}{E_{nk}-H_{0}+\mathrm{i}\epsilon}V\left\vert
\psi_{nk}^{+}\right\rangle.
\end{equation}
Then, the probability amplitude of the GA's excitation reads
\begin{equation}  \label{M-07}
u_{kn}=\frac{\left( J_{1}^{\ast}e^{\mathrm{i}k\frac{d}{2}}+J_{2}^{\ast }e^{-%
\mathrm{i}k\frac{d}{2}}\right) \left( \delta_{n+}\sin\frac{\theta}{2}%
-\delta_{n-}\cos\frac{\theta}{2}\right) }{\Delta_{k}^{n}+\mathrm{i}\Gamma+%
\mathrm{i}\gamma\left( \sin^{2}\frac{\theta}{2}e^{-\mathrm{i}\nu
_{+}\tau}+\cos^{2}\frac{\theta}{2}e^{-\mathrm{i}\nu_{-}\tau}\right) e^{%
\mathrm{i}E_{nk} \tau}}
\end{equation}
with $\tau=d/v$ the time taken by a photon to travel from one coupling point
to the other coupling point, $\Delta_{k}^{n}=v|k|+\nu_{n}-\omega_{e}$ and
\begin{equation}  \label{M-08}
\Gamma =\frac{\pi}{v}\left( |J_{1}|^{2}+|J_{2}|^{2}\right) ,\gamma=\frac{2\pi%
}{v}|J_{1}J_{2}|\cos\varphi_J.
\end{equation}
Here, $\varphi_J=\varphi_1-\varphi_2$ and $\varphi_j$ is the phase of the
coupling strength $J_j$. Note that $\Gamma=\Gamma_{+}+\Gamma_{-}$ is also
the radiative decay rates of a small atom, $\Gamma_{+}=\Gamma\sin^2(\theta/2)
$ and $\Gamma_{-}=\Gamma\cos^2(\theta/2)$ are the spontaneous decay rates of
transitions $|e\rangle\leftrightarrow|\eta_{+}\rangle$ and $%
|e\rangle\leftrightarrow|\eta_{-}\rangle$ due to the coupling with the
waveguide, respectively. In a similar way, the waveguide-mediated nonlocal
damping $\gamma$ can be written as $\gamma=\gamma_{+}+\gamma_{-}$ with $%
\gamma_{+}=\gamma\sin^2(\theta/2)$ and $\gamma_{-}=\gamma\cos^2(\theta/2)$.

In the scattering theory formalism, the scattering matrix with element
\begin{eqnarray}  \label{M-09}
S_{k^{\prime}l;kn}=&-2\pi\mathrm{i}\delta\left(
E_{k^{\prime}l}-E_{kn}\right) \left(\delta_{l+}\sin\frac{\theta}{2}%
-\delta_{l-}\cos\frac{\theta}{2}\right) \\
&\times\left( J_{1}e^{-\mathrm{i}k^{\prime}\frac{d}{2}}+J_{2}e^{\mathrm{i}%
k^{\prime}\frac{d}{2}}\right) u_{kn}+\delta_{nl}\delta\left(
k^{\prime}-k\right).  \notag
\end{eqnarray}
connects states in the asymptotic past with states in the asymptotic future.
The conservation of energy is showed in Eq.(\ref{M-09}) via the delta Dirac
function. For single photons incident from the negative channel with wave
vector $k>0$, the properties of the delta Dirac function give rise to the
scattering amplitudes
\begin{subequations}
\label{M-10}
\begin{eqnarray}
t_{k-} & =\frac{\Delta_{k}^{-}-\frac{2\pi}{v}J_{1c}J_{2c}^{\ast}\sin\left(
\Delta_{k}^{-}\tau+\phi_{-}\right) +\mathrm{i}\Gamma_{+}+\mathrm{i}\gamma
_{+}e^{\mathrm{i}\left( \Delta_{k}^{-}\tau+\phi_{+}\right) }}{\Delta_{k}^{-}+%
\mathrm{i}\left( \Gamma_{+}+\gamma_{+}e^{\mathrm{i}\phi_{+}}e^{\mathrm{i}%
\Delta_{k}^{-}\tau}\right) +\mathrm{i}\left( \Gamma_{-}+\gamma_{-}e^{\mathrm{%
i}\phi_{-}}e^{\mathrm{i}\Delta_{k}^{-}\tau}\right) } \\
r_{k-} & =-\frac{\mathrm{i}\gamma_{-}+\frac{\pi}{v}\mathrm{i}\left[
\left\vert J_{1c}\right\vert ^{2}e^{\mathrm{i}\left( \Delta_{k}^{-}\tau
+\phi_{-}\right) }+\left\vert J_{2c}\right\vert ^{2}e^{-\mathrm{i}\left(
\Delta_{k}^{-}\tau+\phi_{-}\right) }\right] }{\Delta_{k}^{-}+\mathrm{i}%
\left( \Gamma_{+}+\gamma_{+}e^{\mathrm{i}\phi_{+}}e^{\mathrm{i}%
\Delta_{k}^{-}\tau}\right) +\mathrm{i}\left( \Gamma_{-}+\gamma_{-}e^{\mathrm{%
i}\phi_{-}}e^{\mathrm{i}\Delta_{k}^{-}\tau}\right) } \\
t_{k-}^{+} & =\frac{e^{\mathrm{i}\phi}\frac{\pi}{v}\mathrm{i}\left(
J_{1s}+J_{2s}e^{\mathrm{i}\Delta_{k}^{-}\tau}e^{\mathrm{i}\phi_{+}}\right)
\left( J_{1c}^{\ast}+J_{2c}^{\ast}e^{-\mathrm{i}\Delta_{k}^{-}\tau }e^{-%
\mathrm{i}\phi_{-}}\right) }{\Delta_{k}^{-}+\mathrm{i}\left( \Gamma
_{+}+\gamma_{+}e^{\mathrm{i}\phi_{+}}e^{\mathrm{i}\Delta_{k}^{-}\tau}\right)
+\mathrm{i}\left( \Gamma_{-}+\gamma_{-}e^{\mathrm{i}\phi_{-}}e^{\mathrm{i}%
\Delta_{k}^{-}\tau}\right) } \\
r_{k-}^{+} & =\frac{e^{\mathrm{i}\phi}\frac{\pi}{v}\mathrm{i}\left( J_{1s}e^{%
\mathrm{i}\Delta_{k}^{-}\tau}e^{\mathrm{i}\phi_{+}}+J_{2s}\right) \left(
J_{1c}^{\ast}+J_{2c}^{\ast}e^{-\mathrm{i}\Delta_{k}^{-}\tau }e^{-\mathrm{i}%
\phi_{-}}\right) }{\Delta_{k}^{-}+\mathrm{i}\left( \Gamma _{+}+\gamma_{+}e^{%
\mathrm{i}\phi_{+}}e^{\mathrm{i}\Delta_{k}^{-}\tau}\right) +\mathrm{i}\left(
\Gamma_{-}+\gamma_{-}e^{\mathrm{i}\phi_{-}}e^{\mathrm{i}\Delta_{k}^{-}\tau}%
\right) }
\end{eqnarray}
where $t_{k-}$ and $r_{k-}$ are the amplitudes of the reflected wave and
transmitted wave, $t_{k-}^{+}$ and $r_{k-}^{+}$ are the forward and backward
transfer amplitudes. Phases in Eq.(\ref{M-10}) are defined as
\end{subequations}
\begin{equation}  \label{M-11}
\phi_{\pm}=\left(\omega_{e}-\nu_{\pm}\right)\tau,\phi=\left(\nu_{+}-\nu_{-}%
\right)\tau/2
\end{equation}
The elastic scattering by the GA preserves the original frequency of single
photons, which is described by the transmittance and reflectance, $%
T=|t_{k-}|^2$ and $R=|r_{k-}|^2$ respectively. The inelastic scattering
changes the frequency of propagating photons by the amount $|\nu_{+}-\nu_{-}|
$, which can be adjusted by the driving field. The probability for photons
with frequency conversion is described by the conversion probability $%
T_c=|r_{k-}^{+}|^2+|t_{k-}^{+}|^2$. To convert frequency, $\theta\neq n\pi$
is required, i.e., the classical field must be applied. $T+R+T_c=1$ for all
frequencies indicates probability conservation for the photon. When the
dissipation of the excited state to the environment except the 1D waveguide
is considered, the total scattering probability is smaller than unity by
reducing the scattering probabilities and increasing the linewidth of the
spectra, so the dissipation of the GA is not taken into account in this
paper. Since $|r_{k-}^{+}|\neq|t_{k-}^{+}|$ as long as the phase $%
\varphi_J\neq n\pi$, the outgoing photon displays a preferred direction,
which is a consequence of breaking time-reversal symmetry. We emphasize that
there is NO chiral coupling of atomic transitions to the waveguide here. The
symmetry breaking should also be shown in the photon scattering process. In
the 1D, its vector nature of the wave vector is reflected in the occurrence
of both positive and negative values. For photons incident from different
direction of the negative channel with wave vector $-k<0$, the element of
the scattering matrix has the same expression as Eq.(\ref{M-09}) with $-k$
replacing $k$, which leads to the following scattering amplitudes
\begin{subequations}
\label{M-12}
\begin{eqnarray}
\tilde{t}_{k-} & =\frac{\Delta_{k}^{-}-\frac{2\pi}{v}J_{1c}^{\ast}J_{2c}\sin%
\left( \Delta_{k}^{-}\tau+\phi_{-}\right) +\mathrm{i}\Gamma _{+}+\mathrm{i}%
\gamma_{+}e^{\mathrm{i}\left( \Delta_{k}^{-}\tau+\phi _{+}\right) }}{%
\Delta_{k}^{-}+\mathrm{i}\left( \Gamma_{+}+\gamma _{+}e^{\mathrm{i}%
\phi_{+}}e^{\mathrm{i}\Delta_{k}^{-}\tau}\right) +\mathrm{i}\left(
\Gamma_{-}+\gamma_{-}e^{\mathrm{i}\phi_{-}}e^{\mathrm{i}\Delta_{k}^{-}\tau}%
\right) } \\
\tilde{r}_{k-} & =-\frac{\mathrm{i}\gamma_{-}+\frac{\pi}{v}\mathrm{i}\left[
\left\vert J_{1c}\right\vert ^{2}e^{-\mathrm{i}\left( \Delta_{k}^{-}\tau
+\phi_{-}\right) }+\left\vert J_{2c}\right\vert ^{2}e^{\mathrm{i}\left(
\Delta_{k}^{-}\tau+\phi_{-}\right) }\right] }{\Delta_{k}^{-}+\mathrm{i}%
\left( \Gamma_{+}+\gamma_{+}e^{\mathrm{i}\phi_{+}}e^{\mathrm{i}%
\Delta_{k}^{-}\tau}\right) +\mathrm{i}\left( \Gamma_{-}+\gamma_{-}e^{\mathrm{%
i}\phi_{-}}e^{\mathrm{i}\Delta_{k}^{-}\tau}\right) } \\
\tilde{t}_{k-}^{+} & =\frac{e^{-\mathrm{i}\phi}\frac{\pi\mathrm{i}}{v}\left[
J_{1s}+J_{2s}e^{-\mathrm{i}\left( \Delta_{k}^{-}\tau+\phi_{+}\right) }\right]
\left( J_{1c}^{\ast}+J_{2c}^{\ast}e^{\mathrm{i}\Delta_{k}\tau }e^{\mathrm{i}%
\phi_{-}}\right) }{\Delta_{k}^{-}+\mathrm{i}\left( \Gamma _{+}+\gamma_{+}e^{%
\mathrm{i}\phi_{+}}e^{\mathrm{i}\Delta_{k}^{-}\tau}\right) +\mathrm{i}\left(
\Gamma_{-}+\gamma_{-}e^{\mathrm{i}\phi_{-}}e^{\mathrm{i}\Delta_{k}^{-}\tau}%
\right) } \\
\tilde{r}_{k-}^{+} & =\frac{e^{-\mathrm{i}\phi}\frac{\pi\mathrm{i}}{v}\left[
J_{1s}e^{-\mathrm{i}\left( \Delta_{k}^{-}\tau+\phi_{+}\right) }+J_{2s}\right]
\left( J_{1c}^{\ast}+J_{2c}^{\ast}e^{\mathrm{i}\Delta_{k}\tau}e^{\mathrm{i}%
\phi_{-}}\right) }{\Delta_{k}^{-}+\mathrm{i}\left( \Gamma_{+}+\gamma_{+}e^{%
\mathrm{i}\phi_{+}}e^{\mathrm{i}\Delta_{k}^{-}\tau }\right) +\mathrm{i}%
\left( \Gamma_{-}+\gamma_{-}e^{\mathrm{i}\phi_{-}}e^{\mathrm{i}%
\Delta_{k}^{-}\tau}\right) }
\end{eqnarray}
Similarly, we define $\tilde{T}=|\tilde{t}_{k-}|^2$, $\tilde{R}=|\tilde{r}%
_{k-}|^2$ and $\tilde{T}_c=|\tilde{r}_{k-}^{+}|^2+|\tilde{t}_{k-}^{+}|^2$ as
the transmittance, reflectance and the conversion probability for incident
photons with $-k$, respectively. It can be found from Eqs.(\ref{M-11}) and (%
\ref{M-12}) that the reflectance $R=\tilde{R}$ and the system presents
reciprocity in transmission and transfer upon reversing the propagation of
an incident photon when $\varphi_J=n\pi$.
\begin{figure}[tbp]
\includegraphics[width=9cm, height=8cm]{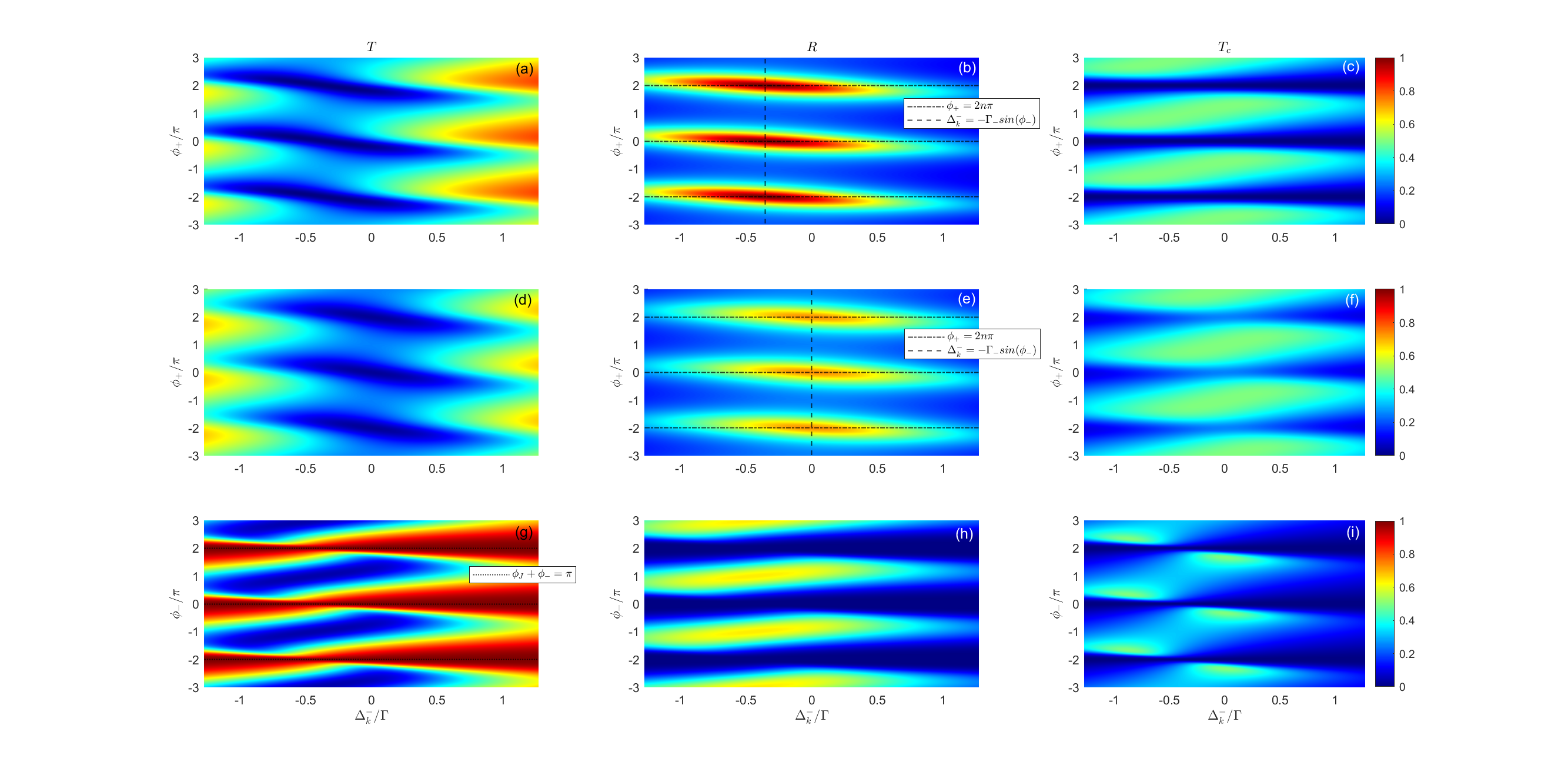}
\caption{(a), (d)/(g) transmittance $T$, (b), (e)/(h) reflectance $R$, and
(c), and (f)/(i) conversion probability $T_c$ versus detuning coefficient $%
\Delta_{k}^{-}/\Gamma$ and phase delay coefficient $(\protect\phi_{+}/%
\protect\pi)$/$(\protect\phi_{-}/\protect\pi)$ for (a)-(c) $\protect\varphi%
_J=\protect\pi, \protect\phi_{-}=0.75\protect\pi$, (d)-(f) $\protect\varphi%
_J=0.75\protect\pi, \protect\phi_{-}=\protect\pi$, and (g)-(i) $\protect%
\varphi_J=\protect\pi, \protect\phi_{+}=\protect\pi/3$. The dashed and
dash-dotted lines in (b) and (e) represent the position for maximum
reflection, respectively. The dash-dotted line in (g) represents the
position for total transmission. We have set $|J_1|=|J_2|$ and $\protect%
\theta=\protect\pi/2$ in all panels.}
\label{fig1}
\end{figure}



\section{\label{Sec:3}Single-photon scattering spectra}

The scattering spectra are determined by the following parameters: the
characteristic frequencies $\omega_e-\nu_{\pm}$, the delay time proportional
to the distance between the two coupling points, and the decay rate $\Gamma$%
. We will set $|J_1|=|J_2|$ since the system reduces to a two-level GA
interaction with a 1D waveguide and total transmission can be found for a
incident single photon when the driving field is absent (i.e., $\theta=0$).
In a GA-waveguide system, $\tau^{-1}, \Gamma<\omega_e-\nu_{\pm}$ is usually
satisfied. So we divided the discussion into two parts depending on whether $%
\tau^{-1}$ is comparable to or smaller than $\Gamma$.


\subsection{\label{Sec:3a}Markovian regime}

When the travel time of light between coupling points is smaller than the
characteristic timescale $\Gamma^{-1}$, the factors $\exp\left({\pm}\mathrm{i%
}\Delta_{k}^{-}\tau\right)$ can be set to 1, but it is impossible to assume
that factor $\exp\left({\pm}\mathrm{i}\phi_{\pm}\right)=1$ since the
distance is at least of the order of a resonant wavelength. Then, the
scattering amplitudes for single photons incident from the negative channel
with wave vector $k>0$ reads
\end{subequations}
\begin{subequations}
\label{SP-01}
\begin{eqnarray}
t_{k-} & =\frac{\Delta_{k}^{-}-\frac{2\pi}{v}J_{1c}J_{2c}^{\ast}\sin\phi
_{-}+\mathrm{i}(\Gamma_{+}+\gamma_{+}e^{\mathrm{i}\phi_{+}})}{\Delta_{k}^{-}+%
\mathrm{i}\left( \Gamma_{+}+\gamma_{+}e^{\mathrm{i}\phi_{+}}\right) +\mathrm{%
i}\left( \Gamma_{-}+\gamma_{-}e^{\mathrm{i}\phi_{-}}\right) } \\
r_{k-} & =-\frac{\mathrm{i}\gamma_{-}+\frac{\pi}{v}\mathrm{i}\left(
\left\vert J_{1c}\right\vert ^{2}e^{\mathrm{i}\phi_{-} }+\left\vert J_{2c}
\right\vert ^{2}e^{-\mathrm{i}\phi_{-} }\right) }{\Delta_{k}^{-}+\mathrm{i}
\left( \Gamma_{+}+\gamma_{+}e^{\mathrm{i}\phi_{+}}\right) +\mathrm{i} \left(
\Gamma_{-}+\gamma_{-}e^{\mathrm{i}\phi_{-}}\right) } \\
t_{k-}^{+} & =\frac{e^{\mathrm{i}\phi}\frac{\pi}{v}\mathrm{i}\left(
J_{1s}+J_{2s}e^{\mathrm{i}\phi_{+}}\right) \left(
J_{1c}^{\ast}+J_{2c}^{\ast}e^{-\mathrm{i}\phi_{-}}\right) }{\Delta_{k}^{-}+%
\mathrm{i}\left( \Gamma_{+}+\gamma_{+}e^{\mathrm{i}\phi_{+}}\right) +\mathrm{%
i}\left( \Gamma_{-}+\gamma_{-}e^{\mathrm{i}\phi_{-}}\right) } \\
r_{k-}^{+} & =\frac{e^{\mathrm{i}\phi}\frac{\pi}{v}\mathrm{i}\left( J_{1s}e^{%
\mathrm{i}\phi_{+}}+J_{2s}\right) \left( J_{1c}^{\ast}+J_{2c}^{\ast}e^{-%
\mathrm{i}\phi_{-}}\right) }{\Delta_{k}^{-}+\mathrm{i}\left(
\Gamma_{+}+\gamma_{+}e^{\mathrm{i}\phi_{+}}\right) +\mathrm{i}\left(
\Gamma_{-}+\gamma_{-}e^{\mathrm{i}\phi_{-}}\right) }
\end{eqnarray}
and the scattering amplitudes for single photons incident from the opposite
direction reads
\end{subequations}
\begin{subequations}
\label{SP-02}
\begin{eqnarray}
\tilde{t}_{k-} & =\frac{\Delta_{k}^{-}-\frac{2\pi}{v}J_{1c}^{\ast}J_{2c}\sin%
\phi_{-}+\mathrm{i}(\Gamma_{+}+\gamma_{+}e^{\mathrm{i}\phi_{+}})}{%
\Delta_{k}^{-}+\mathrm{i}\left( \Gamma_{+}+\gamma_{+}e^{\mathrm{i}%
\phi_{+}}\right) +\mathrm{i}\left( \Gamma_{-}+\gamma_{-}e^{\mathrm{i}%
\phi_{-}}\right) } \\
\tilde{r}_{k-} & =-\frac{\mathrm{i}\gamma_{-}+\frac{\pi}{v}\mathrm{i}\left(
\left\vert J_{1c}\right\vert ^{2}e^{-\mathrm{i}\phi_{-} }+\left\vert
J_{2c}\right\vert ^{2} e^{\mathrm{i}\phi_{-} }\right) }{\Delta_{k}^{-}+%
\mathrm{i}\left( \Gamma_{+}+\gamma_{+}e^{\mathrm{i}\phi_{+}}\right) +\mathrm{%
i}\left( \Gamma_{-}+\gamma_{-}e^{\mathrm{i}\phi_{-}}\right) } \\
\tilde{t}_{k-}^{+} & =\frac{\frac{\pi\mathrm{i}}{v}e^{-\mathrm{i}\phi}\left[
J_{1s}+J_{2s}e^{-\mathrm{i}\phi_{+}}\right] \left(
J_{1c}^{\ast}+J_{2c}^{\ast}e^{\mathrm{i}\phi_{-}}\right) }{\Delta_{k}^{-}+%
\mathrm{i}\left( \Gamma_{+}+\gamma_{+}e^{\mathrm{i}\phi_{+}}\right) +\mathrm{%
i}\left( \Gamma_{-}+\gamma_{-}e^{\mathrm{i}\phi_{-}}\right) } \\
\tilde{r}_{k-}^{+} & =\frac{\frac{\pi\mathrm{i}}{v}e^{-\mathrm{i}\phi}\left[
J_{1s}e^{-\mathrm{i}\phi_{+}}+J_{2s}\right] \left(
J_{1c}^{\ast}+J_{2c}^{\ast}e^{\mathrm{i}\phi_{-}}\right) }{\Delta_{k}^{-}+%
\mathrm{i}\left( \Gamma_{+}+\gamma_{+}e^{\mathrm{i}\phi_{+}}\right) +\mathrm{%
i}\left( \Gamma_{-}+\gamma_{-}e^{\mathrm{i}\phi_{-}}\right) }
\end{eqnarray}
Reciprocal photon transmission can be found from Eqs.(\ref{SP-01}) and (\ref%
{SP-02}) when $\phi_{-}=n\pi$. The Markovian approximation makes the lamb
shift of the excited state from a energy-dependent quantity to a
energy-independent quantity.
\begin{figure}[tbp]
\includegraphics[width=9cm, height=8cm]{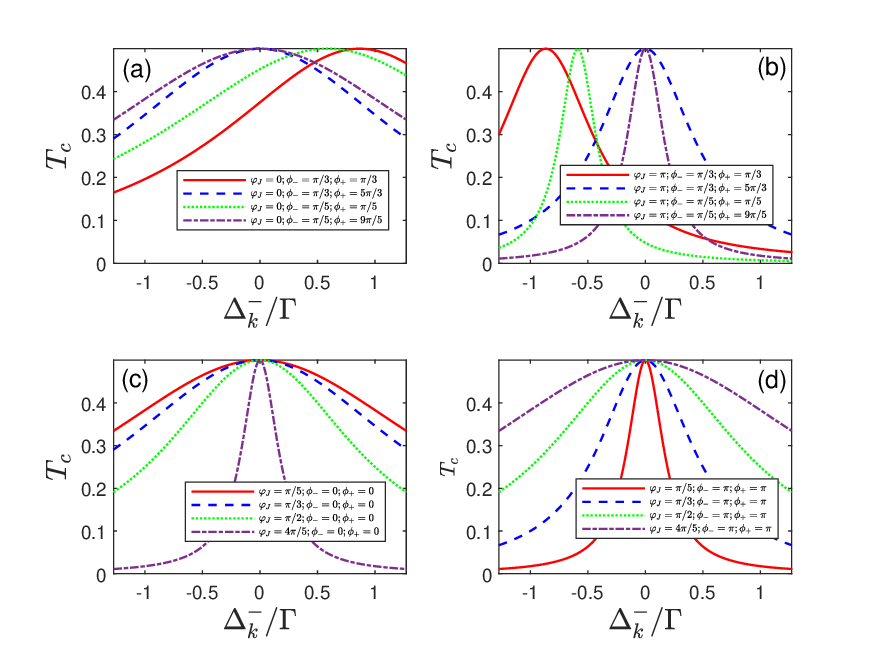}
\caption{The conversion probability $T_c$ versus detuning $%
\Delta_{k}^{-}/\Gamma$ for given $\protect\phi_{\pm},\protect\varphi_J$. We
have set $|J_1|=|J_2|$ and $\protect\theta=\protect\pi/2$ in all panels.}
\label{fig2}
\end{figure}

In Fig.\ref{fig1}, we plot the transmittance $T$, reflectance $R$ and the
conversion probability $T_c$ as functions of the scaled detuning $%
\Delta_{k}^{-}/\Gamma$ and the phase delay $\phi_{+}/\pi$ in the upper and
middle panels, or $\phi_{-}/\pi$ in the lower panels. All scattering
coefficients change periodically with $\phi_{\pm}$. From the upper three
panels (a-c), we found that when $\phi_{+}=2n\pi$, $R=1$ and $T=0$ can be
achieved at $\Delta_{k}^{-}=-\Gamma_{-}\sin\phi_{-}$, and photons undergo
only elastic scattering since $T_c=0$ for all incident photons with
arbitrary frequencies. When we interchange the values of $\varphi_{J}$ and $%
\phi_{-}$, the reflectance $R$ still achieves its maximum value at $%
\Delta_{k}^{-}=0$ and $\phi_{+}=2n\pi$, but its magnitude decreases (see Fig.%
\ref{fig1}e) and frequency conversion happens for photons (see Fig.\ref{fig1}%
f). The lower three panels (g-i) show a frequency-independent total
transmission $T=1$ at $\phi_{-}=2n\pi$. It can be easily found from Eqs.(\ref%
{SP-01}) and (\ref{SP-02}) that the emission from the $|e\rangle%
\leftrightarrow|\eta_{+}\rangle (|\eta_{-}\rangle)$ transition is completely
suppressed at $\varphi_{J}=(2n+1)\pi$ and $\phi_{+}=2n\pi$ ($\phi_{-}=2n\pi$%
) regardless of the value of $\phi_{-}$ ($\phi_{+}$), such suppression is
caused by the formation of a GA-photon bound state. This GA-photon bound
state gives rise to a photonic bound state in continuum (BIC) in the
positive channel and negative channel~\cite{GAB111PRA,GAB111PRA1},
respectively. Hence a total reflection $R=1$ can be observed as long as a
BIC is formed in the positive channel, and a frequency-independent perfect
transmission can be found once a BIC is formed in the negative channel. The
frequency-independent $T_c=0$ occurs when a BIC is formed in either of
channels. With these underlying physics, one can found that the BIC in the
positive (negative) channel also appears at $\varphi_{J}=2n\pi$ and $%
\phi_{+}=(2m+1)\pi$ ($\phi_{-}=(2m+1)\pi$) for arbitrary $\phi_{-}$ ($%
\phi_{+}$). Thus, in addition to $T_c=0$, $R=1$ and $T=0$ can be achieved at
$\Delta_{k}^{-}=\Gamma_{-}\sin\phi_{-}$ for any $\phi_{-}\neq (2m+1)\pi$
when $\varphi_{J}=2n\pi$ and $\phi_{+}=(2n+1)\pi$, the frequency-independent
perfect transmission $T=1$ occurs for arbitrary $\phi_{+}$ when $%
\varphi_{J}=2n\pi$ and $\phi_{-}=(2n+1)\pi$. To enhance the conversion
probability with reciprocity presented, one has to sabotage the condition
for the formation of BICs. In Fig.~\ref{fig2}, we have plotted the $T_c$ as
a function of detuning $\Delta_{k}^{-}/\Gamma$ for given phases, it shows
that in the presence of reciprocity, the optimal conversion probability is
at most, one-half which appears at $\phi_{+}\pm\phi_{-}=2m\pi$ for any $%
\phi_{-}\neq (n+1)\pi$ when $\varphi_{J}=n\pi$ or at $\phi_{+}=2n\pi+%
\phi_{-}=m\pi$ for any $\varphi_{J}\neq n\pi$. The width of the conversion
profile decreases gradually as $\varphi_{J}$ increases from $0$ to $\pi$
when $\phi_{+}=2n\pi+\phi_{-}=2m\pi$ and is reversed when $%
\phi_{+}=2n\pi+\phi_{-}=(2m+1)\pi$, however, the position of its peak is
unchanged during this process, as shown in Fig.~\ref{fig2}(c,d). One
interested condition is $\varphi_{J}=(n+1/2)\pi$ where the collective
damping parameter $\gamma$ induced by the waveguide is zero, $T_c$ is
independent of $\phi_{+}$ (not shown here).

\begin{figure}[tbp]
\includegraphics[width=9cm, height=8cm]{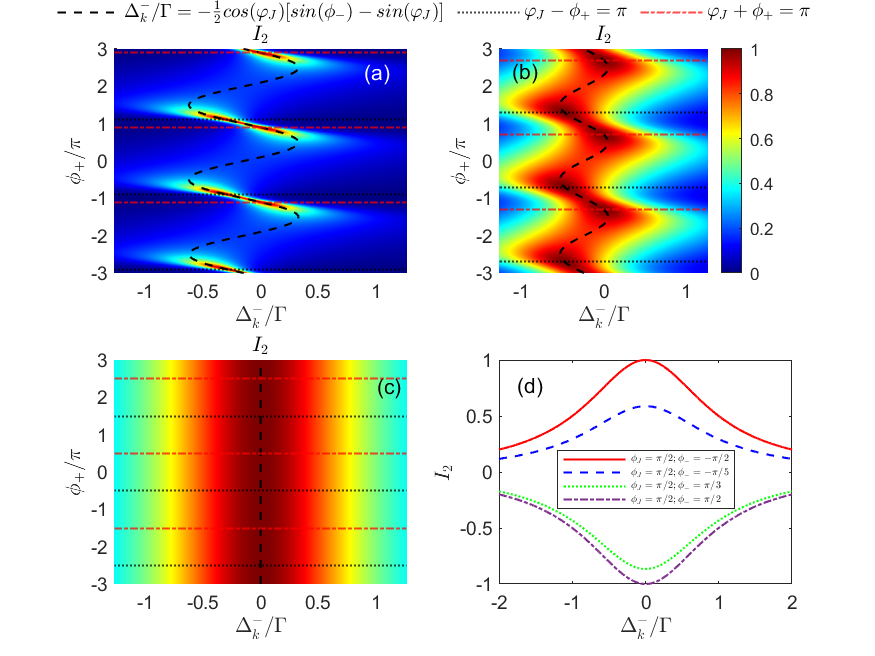}
\caption{The conversion contrast $I_2$ versus detuning $\Delta_{k}^{-}/\Gamma
$ and $\protect\phi_{+}$ for (a) $\protect\varphi_{J}=0.1\protect\pi,
\protect\phi_{-}=1.1\protect\pi$, (b)$\protect\varphi_{J}=0.3\protect\pi,
\protect\phi_{-}=1.3\protect\pi$, (c) $\protect\varphi_{J}=0.5\protect\pi,
\protect\phi_{-}=1.5\protect\pi$. (d) The conversion contrast for different $%
\protect\phi_{-}$ when $\protect\varphi_{J}=\protect\pi/2$. We have set $%
|J_1|=|J_2|$ and $\protect\theta=\protect\pi/2$.}
\label{fig3}
\end{figure}
To increase the conversion probability further (larger than one-half), it is
necessary to break the reciprocity. Eqs.(\ref{SP-01}) and (\ref{SP-02}) told
us that locking either of the relations $\phi_{-}\pm \varphi_{J}=(2n+1)\pi$
could lower the reflectance sine $R=\tilde{R}$. To explore the
unidirectional light transport capabilities of the system, we define the
transmission contrast $I_1=T-\tilde{T}$ and the conversion contrast $I_2=T_c-%
\tilde{T}_c$. The energy conservation in the system indicates $I_1=-I_2$.
The reciprocal scattering corresponds to $I_1=I_2=0$. The optimal
nonreciprocal scattering yields $|I_1|=|I_2|=1$, then the efficiency of the
frequency conversion approaches unity since the scattering coefficients can
not be smaller than zero. We have plotted the conversion contrast $I_2$
versus the scaled detuning $\Delta_{k}^{-}/\Gamma$ and phase delay $\phi_{+}$
with $\phi_{-}-\varphi_{J}=\pi$ in Fig.~\ref{fig3}(a-c). Fig.~\ref{fig3}(a)
and (b) show that $I_2$ can reaches maximum at $\Delta_{k}^{-}=\gamma_{+}%
\sin\phi_{+}+\gamma_{-}\sin\phi_{-}$, its optimal conversion probability is
one, which occurs at two positions (see the cross between the dashed line
and the red dotted-dash line for $\phi_{+}+\varphi_{J}=\pi$ as well as the
cross between the dashed line and the dotted line for $\phi_{+}-\varphi_{J}=%
\pi$). One is located at $\Delta_{k}^{-}=0$, i.e., the incident photon is on
resonance with the $|e\rangle\leftrightarrow|\eta_{-}\rangle$ transition,
the reason for the appearance of $\Delta_{k}^{-}=0$ is that the lamb shift
induced by the two channels cancels out each other due to $%
\phi_{+}+\phi_{-}=2n\pi$. However, the lamb shift presents at the condition $%
\phi_{+}-\phi_{-}=2n\pi$, which results in the other position of the
optimum. Fig.~\ref{fig3}(c) shows that the unity of the conversion contrast
always occurs at $\Delta_{k}^{-}=0$, and it is independent of phase delay $%
\phi_{+}$ when $\varphi_J=(n+0.5)\pi$, i.e., it is only a function of $%
\Delta_{k}^{-}$. So we plot Fig.~\ref{fig3}(d) to present how the optimal $%
I_2$ depends on the phase $\phi_{-}$ at $\varphi_{J}=\pi/2$. It shows that
the phase delay $\phi_{-}$ controls the optimal value of $I_2$ when the
waveguide-mediated nonlocal damping $\gamma=0$.
\begin{figure}[tbp]
\includegraphics[width=9cm, height=8cm]{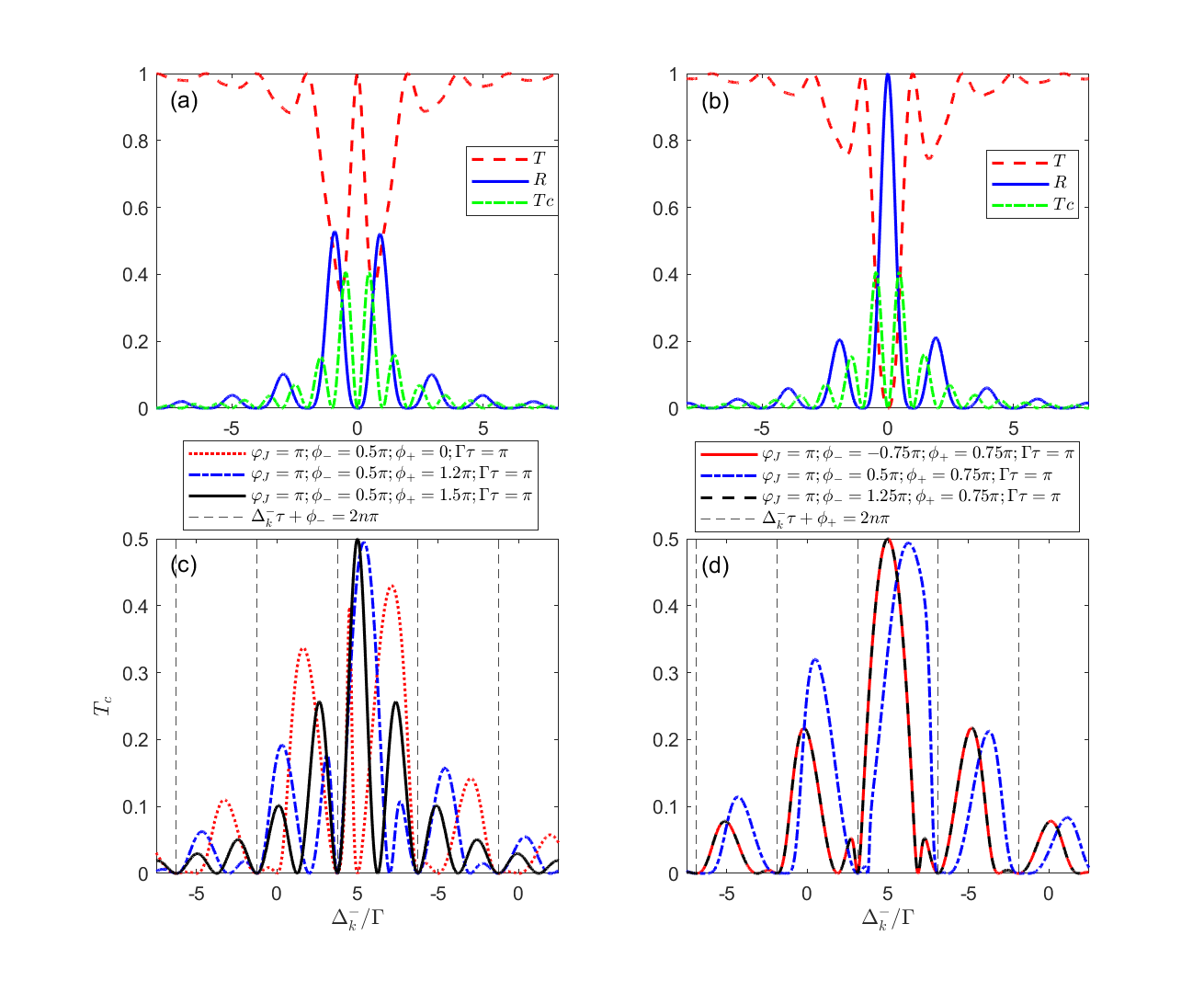}
\caption{(a, b)The transmittance $T=\tilde{T}$ (dashed red curve),
reflectance $R=\tilde{R}$ (solid blue curve) and the conversion probability $%
T_c=\tilde{T}_c$ (dotted-dash green curve) versus detuning $%
\Delta_{k}^{-}/\Gamma$ when $\protect\varphi_{J}=(2n+1)\protect\pi$, (a) $%
\protect\tau\Gamma=1.0025\protect\pi$ and (b) $\protect\tau\Gamma=0.9975%
\protect\pi$. We have set $|J_1|=|J_2|$, $\protect\theta=\protect\pi/2$, $%
\protect\omega_e=600\Gamma$, $\Omega=1.5\Gamma$. (c, d) The conversion
probability $T_c$ versus detuning $\Delta_{k}^{-}$ for given $\protect\phi%
_{\mp}$ and $\protect\varphi_{J}=(2n+1)\protect\pi$.}
\label{fig4}
\end{figure}

\subsection{\label{Sec:3b}Non-Markovian regime}

When the travel time of light between coupling points is comparable to the
characteristic timescale $\Gamma^{-1}$, the factors $\exp\left({\pm}\mathrm{i%
}\Delta_{k}^{-}\tau\right)$ varies with detuning $\Delta_{k}^{-}$, so it is
impossible to keep the reflectance zero by locking the relation of the
phases or to observe the frequency-independent perfect transmission. In Fig.~%
\ref{fig4}(a,b), we have plotted the transmittance, the reflectance and the
conversion probability as a function of the scaled detuning $%
\Delta_{k}^{-}/\Gamma$ when $\varphi_{J}=n\pi$. The spectra in Fig.~\ref%
{fig4}(a,b) show multiple peaks and dips, and the number of peaks and dips
increase with $\tau$ increases. The emission from the $|e\rangle%
\leftrightarrow|\eta_{-}\rangle$ transition is completely suppressed when $%
\Delta_{k}^{-}\tau+\phi_{-}=2m\pi$ with $\varphi_{J}=(2n+1)\pi$, so the
peaks of the transmittance reach unity, and the reflectance and conversion
probability vanish correspondingly. The condition $\Delta_{k}^{-}\tau+%
\phi_{+}=2m\pi$ with $\varphi_{J}=(2n+1)\pi$ leads to the suppression of
emission from the $|e\rangle\leftrightarrow|\eta_{+}\rangle$ transition, so
the conversion probability takes the value of zero. Although there are many
peaks of reflectance, the total reflectance requires the real part of the
denominator in Eqs.(\ref{M-11}) and (\ref{M-12}) to be zero and the
suppression of the emission from the $|e\rangle\leftrightarrow|\eta_{+}%
\rangle$ transition. Note that the positions where the peaks of the
reflectance located do not correspond to those of the conversion dips. In
Fig.~\ref{fig4}(c,d), we plot $T_c$ versus detuning for given $\phi_{\pm}$
and $\varphi_{J}=(2n+1)\pi$. The conversion dips can be divided into two
categories: one is static, while the other moves with the phase $\phi_{+}$
in Fig.~\ref{fig4}c ($\phi_{-}$ in Fig.~\ref{fig4}d). The static dips
satisfy $\Delta_{k}^{-}\tau+\phi_{-}=2m\pi$ in Fig.~\ref{fig4}c ($%
\Delta_{k}^{-}\tau+\phi_{+}=2n\pi$ in Fig.~\ref{fig4}d), the moving dips
satisfy $\Delta_{k}^{-}\tau+\phi_{+}=2n\pi$ in Fig.~\ref{fig4}c ($%
\Delta_{k}^{-}\tau+\phi_{-}=2n\pi$ in Fig.~\ref{fig4}d), it indicates that
the dips of $T_c$ arise from the suppression of emission in either positive
channel or negative channel. Thence, one can control the positions of the
conversion dips flexibly by tuning the phases $\phi_{\pm}$. Fig.~\ref{fig4}%
(c,d) show that the conversion probability is symmetric to $\Delta_{k}^{-}=0$
when $\phi_{+}+\phi_{-}=2n\pi$, see the black solid line in Fig.~\ref{fig4}%
(c) as well as the black dashed line and the red solid line in Fig.~\ref%
{fig4}(d); the optimal conversion probability is at most one-half since
photon scattering coefficients are symmetric for the forward and backward
propagation directions. Thus as long as the system is reciprocal, one-half
is the upper bound of the conversion probability.
\begin{figure}[tbp]
\includegraphics[width=9cm, height=8cm]{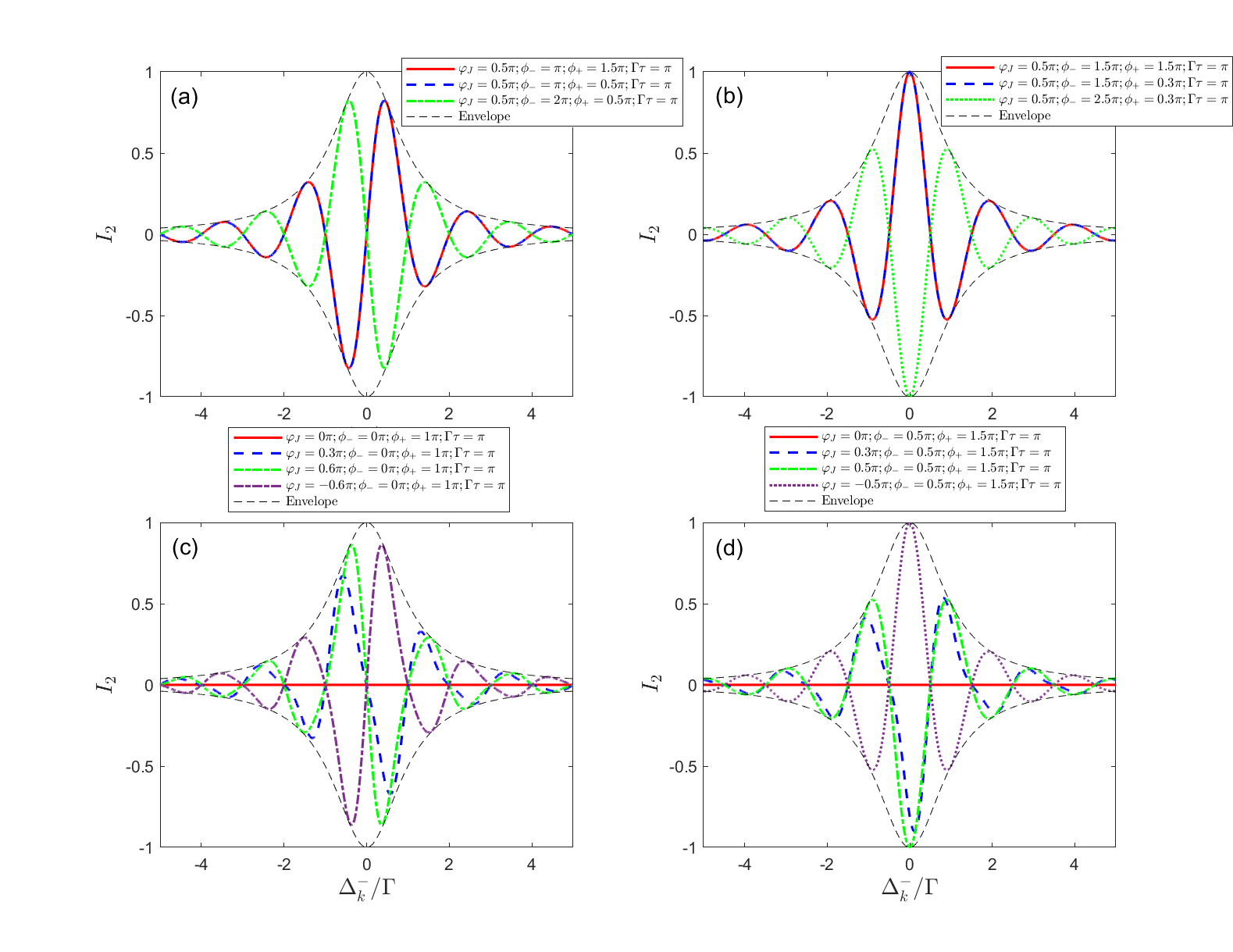}
\caption{The conversion contrast $I_2$ versus detuning $\Delta_{k}^{-}/\Gamma
$ when $\Gamma\protect\tau=\protect\pi$. (a,b) $\protect\varphi_{J}=0.5%
\protect\pi$, (c) $\protect\phi_{-}=0, \protect\phi_{+}=\protect\pi$ (d) $%
\protect\phi_{-}=0.5\protect\pi, \protect\phi_{+}=1.5\protect\pi$.}
\label{fig5}
\end{figure}

The previous discussion demonstrates that preventing the suppression of the
GA's emission is the first step to improving conversion probability, and
breaking the reciprocity is the second step to further enhance the
conversion probability. When $\varphi_{J}=\pi/2$, the nonlocal damping $%
\gamma$ is absent, the GA in the excited state decays into states $%
|\eta_{-}\rangle$ and $|\eta_{+}\rangle$ at the same rate, and the Lamb
shift induced by the vacuum disappears. Fig.~\ref{fig5}(a,b) depicts the
conversion contrast $I_2$ versus the detuning when $\Gamma\tau=\pi$. Similar
to $I_2$ in the Markovian regime, the conversion contrast $I_2$ is also
independent of phase delay $\phi_{+}$ (see the red solid line and the blue
dashed line in Fig.~\ref{fig5}(a,b)), and phase delay $\phi_{-}$ controls
its optimal value. We also found that the decay rate of the GA to both
channels are equal and the Lamb shift is absent when $\phi_{+}-%
\phi_{-}=(2n+1)\pi$, $I_2$ in this case is also plotted in Fig.~\ref{fig5}%
(c,d) for fixed $\phi_{\pm}$ and different $\varphi_{J}$. The graphs of Fig.~%
\ref{fig5} exhibit oscillatory behaviors, and the peaks and dips of these
oscillations form stable envelopes. Graphs in Fig.~\ref{fig5}(a,c) are
symmetric about the point $\Delta_{k}=0$, i.e., $I_2$ is an odd function of $%
\Delta_{k}$. Though one can improve $I_2$ closer to one by increasing $\tau$%
, the conversion contrast of one is unattainable. Graphs in Fig.~\ref{fig5}%
(b,d) are symmetric about the $\Delta_{k}^{-}=0$ axis, i.e., $I_2$ is an
even function of $\Delta_{k}$, thus the conversion contrast reaches $\pm1$.
The conversion probability $T_c=1$ when $I_2=1$. Besides the vanishing of
the reflectance, the waves emitted by the GA and propagated along the
forward direction in the waveguide interference destructively for photon
incident from the forward direction, however, the waves emitted by the GA
and propagated along the backward direction in the waveguide interference
constructively for photon incident from the backward direction. In addition,
Fig.~\ref{fig5}(c,d) show that the optimal nonreciprocal scattering only
achieves at $\varphi_{J}=(n+0.5)\pi$.


\section{\label{Sec:5}Conclusion}

We consider a V-type GA, where the field in a conventional waveguide with
linear light dispersion and the driving field interact with the GA on the
transitions $|g\rangle\leftrightarrow|e\rangle$ and $|g\rangle%
\leftrightarrow|f\rangle$, correspondingly. Single photons traveling in the
waveguide can be scattered by the GA elastically and inelastically at two
coupling points. The inelastic scattering realizes the single-photon
frequency conversion as the GA spontaneously decays to a superposition state
of $|g\rangle$ and $|f\rangle$ orthogonal to its initial state. The
scattering coefficients for photons incident from forward and backward
directions are obtained via the Lippmann-Schwinger equation, which display
quite different scattering spectra in the Markovian and Non-Markovian
regimes which is defined according to the relative relation of the travel
time of the photon between the coupling points and the inverse of the bare
relaxation rate of the GA. We have studied the influence of the accumulated
phases $\phi_{\pm}$ that photons travel between coupling points and the
phase difference $\varphi_j$ between the two coupling strengths on the
scattering spectra as well as the conversion contrast which characterize
nonreciprocal single-photon scattering in both regimes. Light transport is
symmetric for the forward and backward propagation direction at $%
\varphi_j=n\pi$ in both regimes, the optimal conversion probability is at
most, one-half, and it can surpass the limit one-half and reach unity by
breaking the reciprocity. In the Markovian regime, in addition to the
reciprocal behavior of single photons at $\phi_{-}=n\pi$, the formation of
the photonic BIC in the elastic scattering process gives rise to the
frequency-independent total transmission, and the BIC in inelastic
scattering process generates the total reflection. In the non-Markovian
regime, the spectra and the contrast show multiple peaks and dips, and the
number of peaks and dips increase as the scale of a two-legged giant atom
increases. Total transmissions occurs only for the single-photon with
frequency which can suppress the emission from the GA's transition to its
initial state. More stringent conditions are required to achieve the total
reflection. The suppression of the emission from either transitions results
in the conversion probability vanishing. When the GA in the excited state
decays into its initial state and the superposition state orthogonal to its
initial state at the same rate, the conversion contrast exhibits the
oscillatory behaviors, the peaks and dips of these oscillations form stable
envelopes, and the conversion contrast can be an odd or even function of the
detuning. Although single-photon frequency up- or down-conversion with
efficiency close to unity can be achieved by increasing the scale of the GA,
its optimal unity occurs only when the incident photon is on resonance with
the GA.

\begin{acknowledgments}
This work was supported by NSFC Grants No.12247105, No.12421005, XJ-Lab Key Project (23XJ02001),
and the Science $\And $ Technology Department of Hunan Provincial Program (2023ZJ1010).
\end{acknowledgments}

\end{subequations}

\end{document}